\documentclass[3p,times]{elsarticle}

\usepackage{ecrc}

\volume{00}

\firstpage{1}

\journalname{Nuclear Physics A}

\runauth{}

\jid{nupha}

\usepackage{amssymb}

\usepackage[figuresright]{rotating}

\begin{document}

\begin{frontmatter}


\title{Measurement of electrons from Heavy Quarks in PHENIX}




\author{Sourav Tarafdar (for PHENIX collaboration)}
\ead{souravbhu@yahoo.com}
\address{Dept. of Physics, Banaras Hindu University, Varanasi, India 221005}

\begin{abstract}
Heavy quarks are one of the most valuable probes for the matter produced in relativistic heavy ion collisions. PHENIX experiment $@$ RHIC is designed specifically to study leptons, so electrons from the decay of heavy quarks are one of the most important tools in PHENIX for the study of heavy flavour production. Measurements of electron spectra at mid rapidity region has been done by PHENIX in broad $p_{T}$ regions for various collision species starting from p+p to Au+Au.

The techniques used by PHENIX for measuring electrons from heavy quarks and the latest results in collision species p+p, d+Au, Cu+Cu and Au+Au will be discussed in this article.
\end{abstract}

\begin{keyword}
heavy ion collision \sep PHENIX \sep heavy quark

\end{keyword}

\end{frontmatter}

\section{Introduction}
\label{intro}
 Measurement of electrons from heavy quarks by PHENIX has been done for various colliding systems and energies. PHENIX has measured the single electron spectra from heavy flavour decays in Au+Au collisions at $\sqrt{S_{NN}} =200$ GeV~\cite{ref1}. The measurement of heavy flavour production for p+p in PHENIX has also been measured~\cite{ref2} which provides the baseline for studying hot and dense matter effects in heavy ion reactions. In Au+Au we observe a strong suppression of single electrons compared with scaled p+p collisons. This evidence is a challenge for radiative energy loss models because heavy quarks are expected to radiate less than light quarks. Measurements done in Cu+Cu collisions by PHENIX provide a measurement for species of intermediate size between pp and Au+Au, while measurements in Au+Au collisions at $\sqrt{S_{NN}} = 62.4$ GeV provide a measurement at lower energy.

\section{Methodology}
\label{method}
The first step of extracting heavy flavour single electrons is to measure electrons from all sources. Electrons are identified in PHENIX using the Ring Imaging Cherenkov counter and Electromagnetic calorimeter. Background estimation and subtraction plays a crucial role in the measurement of single electron from heavy flavours. The background is estimated by two independent approaches~\cite{ref3} :
\begin{itemize}
\item Estimation of photonic and non photonic background electrons using PHENIX hadron decay generator.
\item Using converter method in which photon converter is added to increase the conversion which increases the number of conversions. Thus one can estimate the conversion generated background from this enriched sample.
\end{itemize}
 The photonic background calculated using the hadron decay generator is scaled to the photonic background from the converter method. Then the combined non photonic and scaled photonic electrons are subtracted from the total (identified) electron yield. The resulting yields from p+p and nuclecus-nucleus systems are then used to derive the nuclear modification factors of the different nuclear species. The $R_{AA}$ is defined as
\begin{eqnarray}
R_{AA} = \frac{dN^{e}_{AA}/dp_{T}}{<N_{coll}>dN^{e}_{pp}/dp_{T}}.
\end{eqnarray} 
where, $dN^{e}_{AA}/dp_{T}$  is the yield in either Au+Au, Cu+Cu or d+Au, $dN^{e}_{pp}/dp_{T}$ being the yield in pp collisons and $N_{coll}$ is the number of nucleon-nucleon collisions in the specific nucleus-nucleus system. $R_{AA}$ is unity under the absence of nuclear effects whereas $R_{AA} \neq 1$ is expected for the presence of nuclear effects. 

\begin{figure}[t] 
\vspace{-3mm}
\begin{center} 
\includegraphics[height=2.5in,width=3.0in]{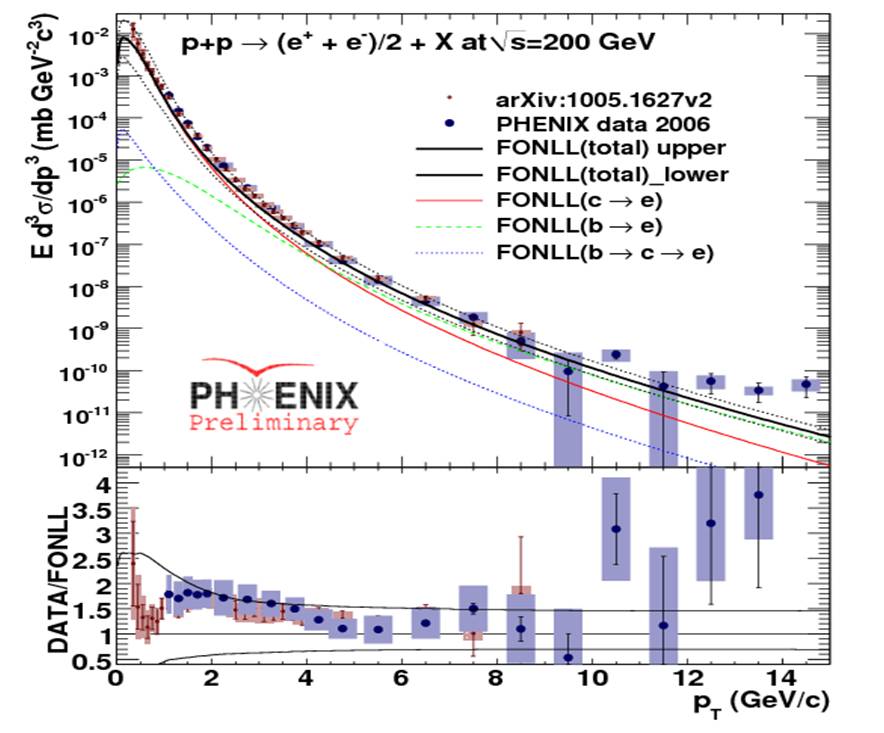}
\includegraphics[height=2.5in,width=3.0in]{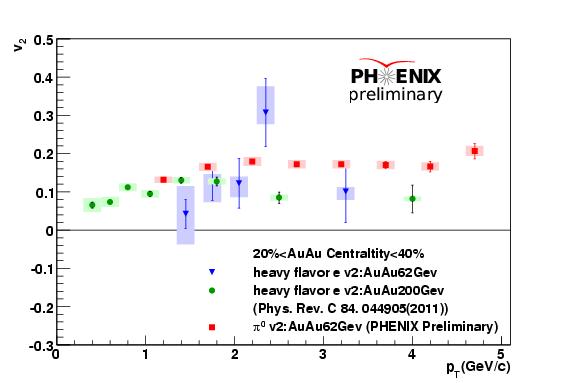} \\
\end{center} 
\vspace{-3mm}
\caption{\small Left panel shown the heavy flavour single electron invariant yield from combined 2005 and 2006 pp data at $\sqrt{S_{NN}} = 200$ GeV. Right panel shown the single electron $v_{2}$ from heavy flavour in Au+Au at $\sqrt{S_{NN}} =  62.4$ GeV. }
\label{fig:singe_v2} 
\end{figure}  
 Other than nuclear modification factor the measurement of azimuthal anisotropy of the particle distribution $v_{2}$ in heavy ion collision is being done which provides us further insight into the energy loss mechanism by heavy quarks.

\section{Results}
\label{results} 
Various results from the measurement of heavy flavour electrons by PHENIX for different collision species has been summarized here.
\subsection{ pp at $\sqrt{S_{NN}} = 200$ GeV }
\label{pp}
As measurements in the pp system acts as a baseline so precision in measurements in pp is an important factor. Recently PHENIX has combined the 2005 and 2006 data providing extension upto high $p_{T}$ and smaller uncertainties. Thus better accuracy of $R_{AA}$ has been attained. Left hand panel of fig.~\ref{fig:singe_v2} shows the invariant yield of single electrons from heavy flavour measured by PHENIX after combining 2005 and 2006 data. The $p_{T}$ spectrum extends upto 15 GeV and has good agreement with FONLL calculations.

\subsection{ Au+Au at $\sqrt{S_{NN}} = 62$ GeV }
Measurement of single electron from heavy quark $v_{2}$ by PHENIX for Au+Au at $\sqrt{S_{NN}} = 200$ GeV showed that the flow reduces from lighter quarks to heavier quarks contradicting the expected behavior.  
Recently measurement of heavy flavour single electron $v_{2}$ for Au+Au at lower beam energy has been done by PHENIX in order to answer the question about flow of heavy quarks at lower beam energy in Au+Au collision. Right hand panel of fig.~\ref{fig:singe_v2} shows the PHENIX preliminary result on heavy flavour single electron $v_{2}$ at $\sqrt{S_{NN}} = 62.4$ GeV. Flow of heavy quarks at lower beam energy in Au+Au collision has been observed which is consistent with $v_{2}$ measurement in Au+Au at $\sqrt{S_{NN}} = 200$ GeV~\cite{ref1}.

\begin{figure}[!t] 
\vspace{-3mm}
\begin{center} 
\includegraphics[height=2.5in,width=3.0in]{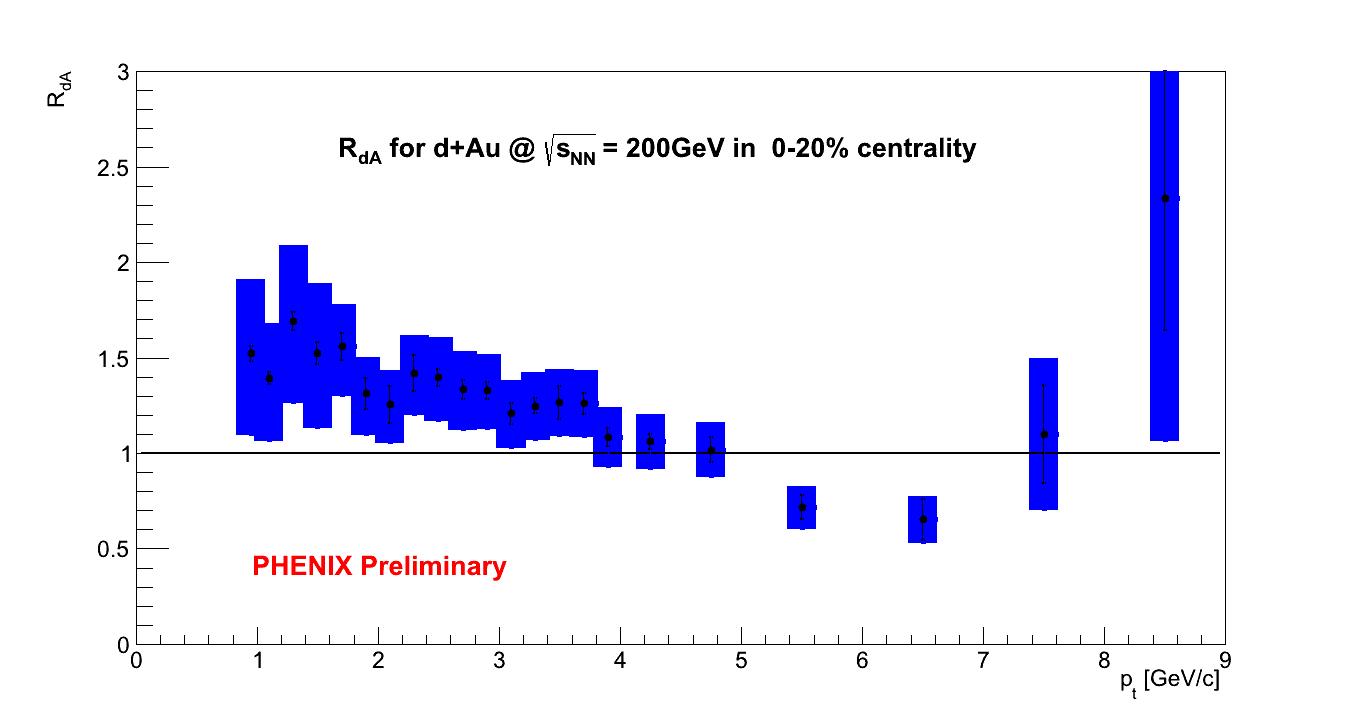}
\includegraphics[height=2.5in,width=3.0in]{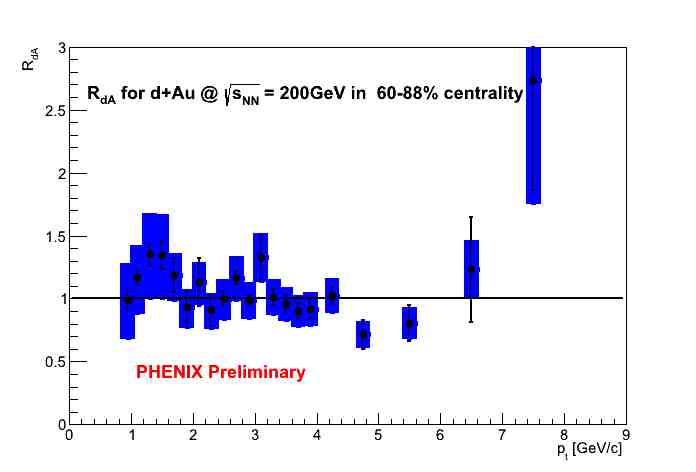} \\
\end{center} 
\vspace{-3mm}
\caption{\small Heavy flavour single electron $R_{dA}$ in most central collisons (left panel) and in most peripheral collisons (right panel) in d+Au at $\sqrt{S_{NN}} = 200$ GeV. }
\label{fig:R_da} 
\end{figure}  

\begin{figure}[!h] 
\vspace{-3mm}
\begin{center} 
\includegraphics[height=2.6in,width=3.0in]{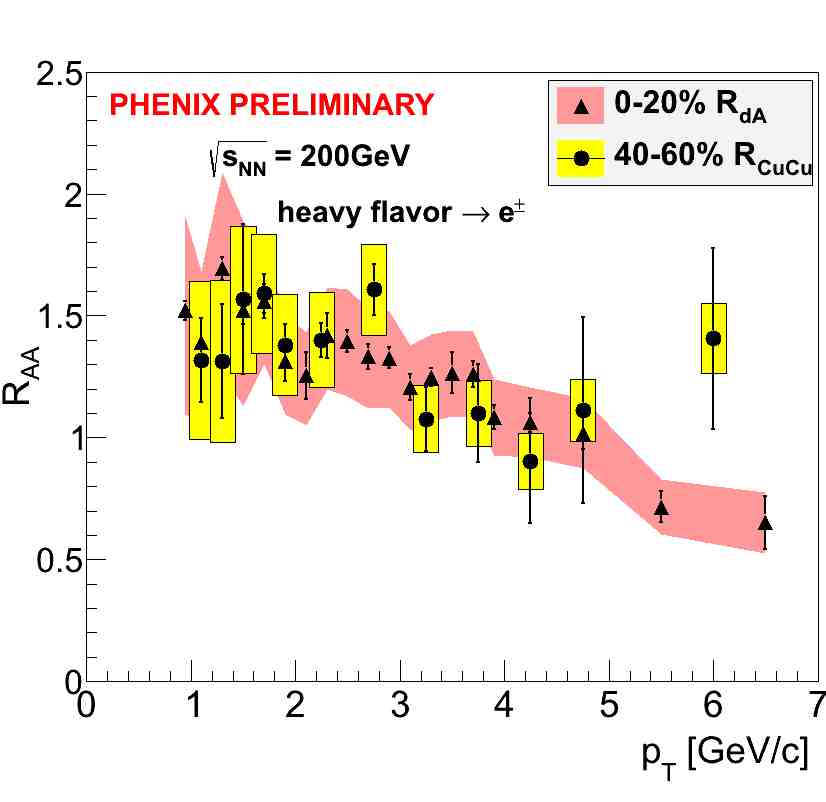}
\includegraphics[height=2.4in,width=3.0in]{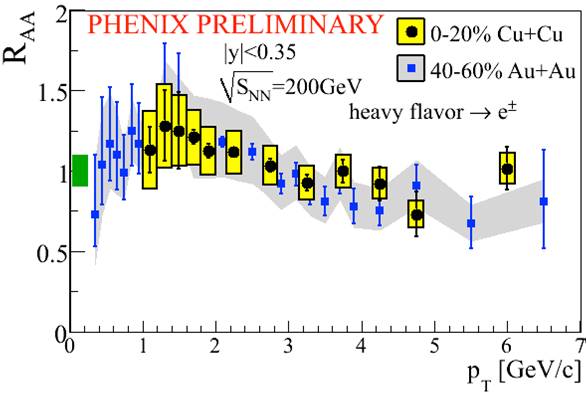} \\
\end{center} 
\vspace{-3mm}
\caption{\small Left panel shows the heavy flavour single electron $R_{CuCu}$ in mid peripheral collisons is consistent with $R_{dA}$ in most central collisions. Right panel shows $R_{CuCu}$ in most central collisions is consistent with $R_{AA}$ in mid peripheral collisions. }
\label{fig:Rda_Rcc} 
\end{figure}  
\subsection{ d+Au at $\sqrt{S_{NN}} = 200$ GeV}
 The measurement of heavy flavours in d+Au provides us an important tool to investigate cold nuclear matter effects. The measurement of single electrons from heavy flavours in d+Au at $\sqrt{S_{NN}} = 200$ GeV has been done by PHENIX using the 2008 RHIC run. Fig.~\ref{fig:R_da} shows the preliminary result of $R_{dA}$ in most central and most peripheral bins. In most central collisions no suppression in $R_{dA}$, unlike in $R_{AA}$~\cite{ref1}, is observed. Thus the suppresion observed in $R_{AA}$ is interpreted as due to the hot and dense medium and is not caused by an initial state effect. In most peripheral collisions $R_{dA} \approx 1$ and hence is consistent with p+p. The results on $R_{dA}$ are the most potential cold nuclear matter effect.

\begin{figure}[t] 
\vspace{-3mm}
\begin{center} 
\includegraphics[height=2.5in,width=3.0in]{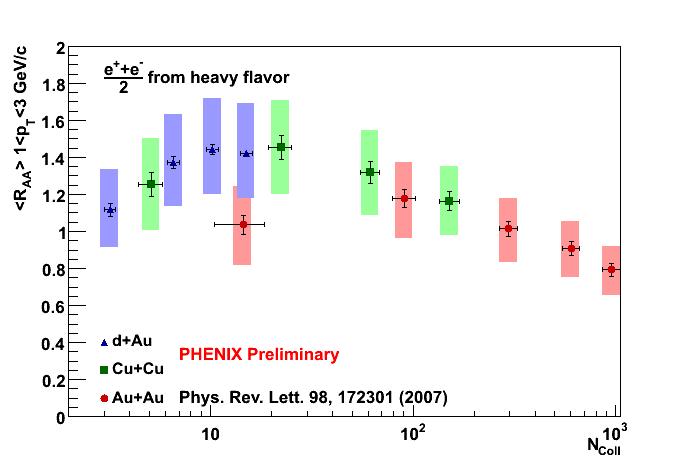}
\includegraphics[height=2.5in,width=3.0in]{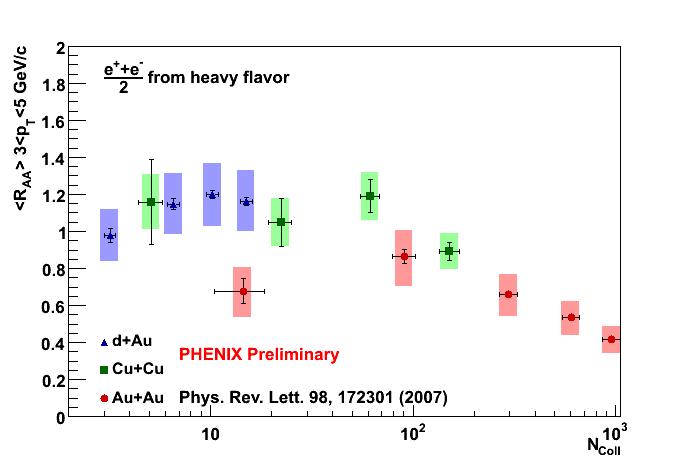} \\
\end{center} 
\vspace{-3mm}
\caption{\small Trend of $R_{CuCu}$, $R_{AuAu}$ and $R_{dA}$ with $N_{coll}$ in $1 < p_{T} < 3 $ GeV/c (left panel) and in $3 < p_{T} < 5 $ GeV/c (right panel). The suppression of heavy flavour single electron from Cu+Cu scales with the one from Au+Au at $N_{coll} \approx$ 100. $R_{CuCu}$ scales with $R_{dA}$ at $N_{coll} \approx$ 15. }
\label{fig:Rda_Rcc_Raa_Ncoll} 
\end{figure}  

\subsection{ Cu+Cu at $\sqrt{S_{NN}} = 200$ GeV}
The measurement of heavy flavour single electrons from Cu+Cu provides important information about the medium produced when the collision species has size between d+Au and Au+Au and is symmetric. Thus it bridges the gap between two different collision systems whose mass and size are at the two extreme points. Left panel of fig.~\ref{fig:Rda_Rcc} shows a preliminary result of $R_{CuCu}$ by PHENIX in mid peripheral collisions which is consistent with $R_{dA}$ in most central ones. Right panel shows $R_{CuCu}$ in most central collisions and is consistent with $R_{AA}$ in the mid peripheral case. Fig.~\ref{fig:Rda_Rcc_Raa_Ncoll} shows the trend of $R_{CuCu}$, $R_{AuAu}$ and $R_{dA}$ with $N_{coll}$. The measurement in $R_{CuCu}$ fills up the mass region between d+Au and Au+Au.
\section{Conclusions}
PHENIX has measured electrons from heavy flavour decay in the mid rapidity region in a wide range of colliding systems and at different collision energies. Preliminary results on p+p, d+Au, Cu+Cu and Au+Au provides us more insight into medium modification. More precise measurement of heavy flavour will be possible with the newly installed silicon vertex tracker by separating charm and bottom.





\bibliographystyle{elsarticle-num}
\bibliography{<your-bib-database>}



\end{document}